# Orbital Quantization in a System of Edge Dirac Fermions in Nanoperforated Graphene

Yu. I. Latyshev[a], A. P. Orlov[a], A. V. Frolov[a,b], V. A. Volkov[a,b], I. V. Zagorodnev[a], V. A. Skuratov[c], Yu. V. Petrov[d], O. F. Vyvenko[d], D. Yu. Ivanov[e], M. Konczykowski[f], and P. Monceau[g]

[a] *Kotel'nikov Institute of Radio Engineering and Electronics, Russian Academy of Sciences, ul. Mokhovaya 11-7, Moscow, 125009 Russia*
[b] *Moscow Institute of Physics and Technology (State University), Institutskii per. 9, Dolgoprudnyi, Moscow region, 141700 Russia*
[c] *Joint Institute for Nuclear Research, Dubna, Moscow region, 141980 Russia*
[d] *Interdisciplinary Resource Center for Nanotechnologies, St. Petersburg State University, St. Petersburg, 198504 Russia*
[e] *Institute of Microelectronic Technology and Ultrahigh-Purity Materials, Russian Academy of Sciences, Chernogolovka, Moscow region, 142432 Russia*
[f] *Ecole Polytechnique, 91128 Palaiseau Cedex, France*
[g] *Institut Neel, CNRS/UJF, UPR2940, 38042 Grenoble, Cedex 9, France*

The dependence of the electric resistance $R$ of nanoperforated graphene samples on the position of the Fermi level $E_F$, which is varied by the gate voltage $V_g$, has been studied. Nanoperforation has been performed by irradiating graphene samples on a Si/SiO$_2$ substrate by heavy (xenon) or light (helium) ions. A series of regular peaks have been revealed on the $R(V_g)$ dependence at low temperatures in zero magnetic field. These peaks are attributed to the passage of $E_F$ through an equidistant ladder of levels formed by orbitally quantized states of edge Dirac fermions rotating around each nanohole. The results are in agreement with the theory of edge states for massless Dirac fermions.

## 1. INTRODUCTION

After a decade since the time of obtaining graphene [1, 2], the investigation of its unusual properties caused by the presence of massless Dirac fermions remains of current interest. The theoretical possibility of the existence of edge states in graphene belongs to such properties [3–5]. One of the first indications of the appearance such states in transport measurements is the observation of Aharonov–Bohm type magnetic oscillations of the resistance in perforated samples of thin graphite [6] and graphene [7] in very high magnetic fields. In the absence of a magnetic field, edge Dirac fermions, if they exist, should move in an effective narrow ring around each nanohole. Since the perimeter of the hole is finite and motion is periodic, the energy of edge Dirac fermions should be orbitally quantized, similar to the energy of the electron in a Bohr atom. The aim of this work is to reveal the orbital quantization levels of edge Dirac fermions in graphene samples with nanoholes by varying the voltage $V_g$ on the control electrode (gate).

## 2. SAMPLES

The resistance of graphene samples on an oxidized highly doped silicon substrate (Si/SiO$_2$) with the thickness $d = 300$ nm of the oxide layer was measured. We studied both graphene samples obtained from Manchester University (Graphene Industries Co) and our samples, which were mechanically exfoliated from natural graphite single crystals using an adhesive tape with the subsequent transfer on the substrate. Nanoholes were created by two methods: first, irradiation by 167-MeV heavy ions (Xe+26) at the ITS-100 cyclotron, Laboratory of Nuclear Reactions, Joint Institute for Nuclear Research, and, second, irradiation by helium ions at the ORION helium ion microscope, St. Petersburg State University.

In the former case, an ensemble of randomly distributed columnar defects is formed. Each defect for electrons in graphene is equivalent to the appearance of one nanohole. The diameter of nanoholes $D$ was estimated with an atomic force microscope (Fig. 1b) (and a scanning electron microscope) as $D \approx 10$ nm. Columnar defects were identified by hillocks of an amorphous material extruded from them [8]. Their average concentration corresponded to a xenon ion fluence of $3 \times 10^9$ cm$^{-2}$. In the latter case, the irradiation of graphene by a helium ion beam with a diameter of 1–2 nm resulted in the formation of a lattice of nanoholes with a diameter of $D \approx 2$ nm and a density of $2 \times 10^{11}$ cm$^{-2}$. In this case, the resistance of the sample increased by a factor of 20 because the mean free path of carriers decreased from 300–500 nm to a value of about the lattice period ($\approx$20 nm). The electric contacts were deposited by means of the laser ablation of gold and had a contact resistance of about 100 Ω.

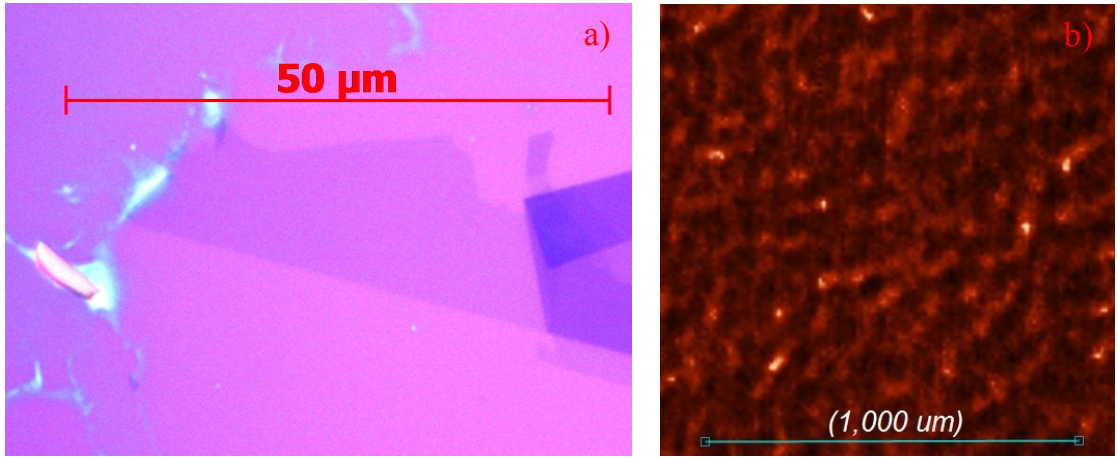

**Fig. 1.** (a) Optical image of the graphene sample on the Si/SiO$_2$ substrate irradiated by heavy ions and (b) image obtained with a scanning atomic force microscope in the phase contrast mode; the scale is 1 μm.

## 3. EXPERIMENT

The measurements were performed in a cryogenic insert with an exchange gas at a fixed temperature in the interval of 1.8–100 K. The graphene samples with the gate electrode were included in a measuring circuit according to the scheme of a field-effect transistor with a common source. The voltage on the gate was varied with a computer-controlled Keithley 2400 SourceMeter at a rate of no higher than 1 V/s with the control of the current through the gate. The resistance of the sample was measured on a weak direct or alternating current (0.1–10 μA). The weakest measuring current was more than three orders of magnitude stronger than the leakage current from the gate into the sample. The dc and ac voltages were measured by a Keithley 2182 nanovoltmeter and an SR 530 two-phase synchronous amplifier. The $R(V_g)$ dependences were usually recorded twice: first, with an increase in $V_g$ and, then, with its decrease at both polarities of $V_g$. The measurements in high magnetic fields were performed at Laboratoires National des Champs Magnetiques Intenses (Grenoble).

Figure 2a shows the $R(V_g)$ dependence for the reference (nonperforated) graphene sample. It has a maximum at the Dirac point, which is shifted toward positive gate voltages by about 6 V; this shift is usually due to the adsorption of water vapor. For this reason, all samples at $V_g = 0$ have a hole conductivity with a hole density of $10^{12}$ cm$^{-2}$ at a mobility of $(1-5) \times 10^3$ cm$^2$/(V s).

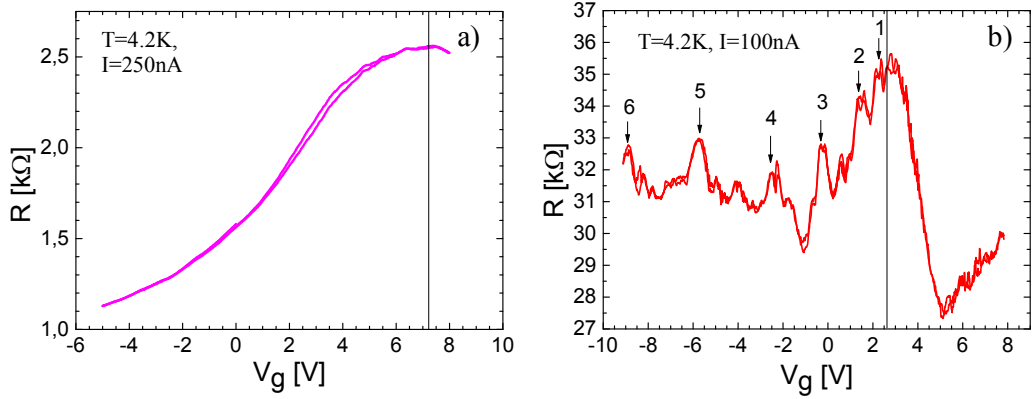

**Fig. 2.** Gate voltage dependence of the electric resistance of the (a) reference graphene sample and (b) graphene sample with columnar defects (sample 3, see Fig. 1b). The vertical straight line marks the voltage corresponding to the Dirac point. The arrows indicate the main series of maxima.

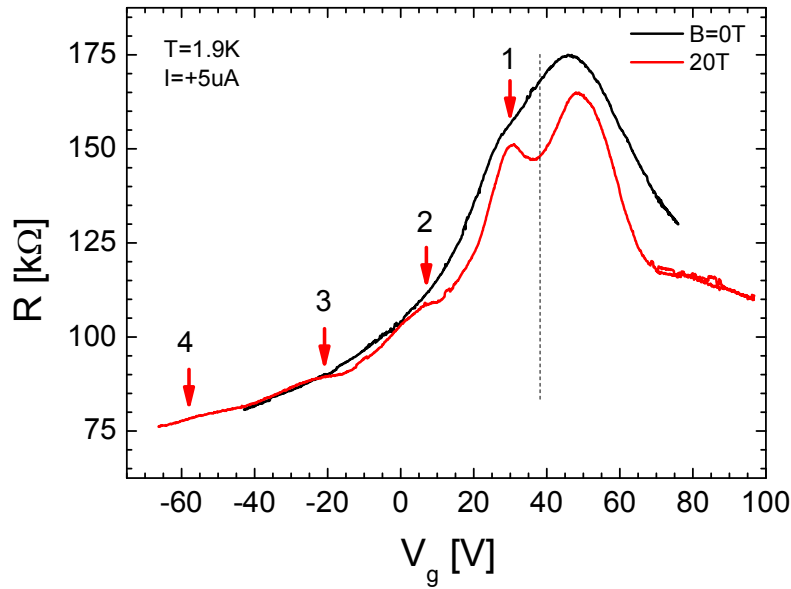

**Fig. 3.** Gate voltage dependence of the resistance of the nanoperforated graphene sample obtained with a helium ion microscope (sample 1). The arrows indicate the positions of peaks in the resistance. The peaks are more pronounced in a magnetic field of 20 T.

A series of peaks appears in the resistance $R(V_g)$ of the graphene samples with columnar defects obtained in Dubna (see Fig. 2b). We discuss only strong peaks (the main series). Similar peaks are also manifested in more "dirty" structures with a lattice of nanoholes obtained at St. Petersburg State University using the helium ion microscope (Fig. 3). The peaks in the latter case are relatively weak. However, they are strongly enhanced in a magnetic field (20 T) at which the magnetic length becomes comparable with the perimeter of a nanohole. A regular series of such peaks in all samples is observed only on the branch of the $R(V_g)$ dependence that corresponds to the hole conductivity. The following universal property is remarkable for the main series of peaks on the $R(V_g)$ curve: the position of the $N$th peak measured from the Dirac point is proportional to $N^2$ for both types of samples (Fig. 4).

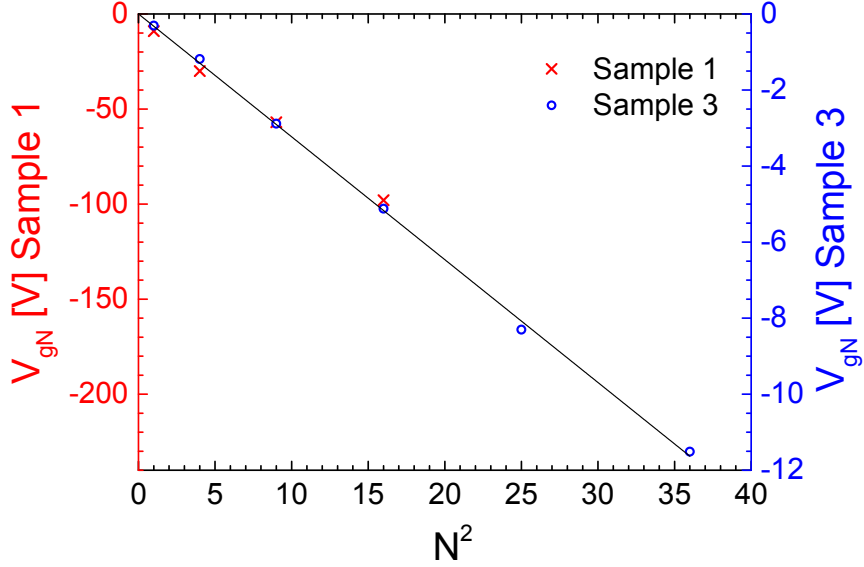

**Fig. 4.** Positions of peaks $V_{gN}$ marked by arrows in Figs. 2 and 3 versus the square of their ordinal number N. The gate voltage $V_g$ was counted from the Dirac point. The data were obtained (circles, right scale) for sample 3 irradiated by heavy ions and (crosses, left scale) for sample 1 irradiated by helium ions. The left and right scales differ by a factor of 20.

## 4. COMPARISON WITH THEORY

The spectrum of surface states for massive Dirac electrons on a half-space was obtained for the first time in [9] (see also review [10]). The boundary of a sample is characterized by the only phenomenological parameter $a$, which appears in the boundary condition for envelope functions and describes the electronic structure of the surface at atomic scales. In the massless limit, the energy of the surface states is related to the two-dimensional tangential component of the momentum $\mathbf{k}_{\parallel}$ as

$$E = 2\hbar a v_F s |\mathbf{k}_{\parallel}| \qquad (1)$$

Here, $v_F$ is the effective "speed of light" in the Dirac equation and the parameter $a$ is for simplicity treated as small ($|a|\ll 1$). The "spin" number $s=\pm 1$ is an eigenvalue of the chirality operator, which is in this case proportional to the mixed products of the vectors of spin, normal, and $\mathbf{k}_{\parallel}$.

In graphene, the valley degree of freedom serves as the spin and the edge of a sample plays the role of the surface. The simplest theory of edge states of massless Dirac fermions in semi-infinite graphene can be developed by neglecting the intervalley interaction. The result at $|a|\ll 1$ is described by Eq. (1) up to the notation: the branches of the one-dimensional tangential momentum $k_{\parallel} > 0$ and $k_{\parallel} < 0$ correspond to different valley quantum numbers $s=\pm 1$ [5, 11, 12]. The microscopic calculation of the boundary (edge in this case) parameter $a$ is a very difficult, practically ill posed problem. We only note that the finiteness of the parameter $a$ results in the asymmetry of spectrum (1). This can be understood in the "inverse heterojunction" model [13]. In this model, the asymmetry of the spectrum appears with allowance for change in the work function on the junction (see references in [10]). We determine the parameter $a$ from the comparison with experiments.

Edge Dirac fermions move along the linear edge at the velocity $v_{edge} = 2av_F$, which is much lower than the Fermi velocity $|a| \ll 1$. In the case of a nanohole, the edge is closed and edge carriers rotate around the hole clockwise or counterclockwise (depending on the number of a valley). We assume that the edge is uniform; i.e., the parameter $a$ remains unchanged in the process of motion along the edge. Then, owing to the orbital quantization of the tangential motion, spectrum (1) becomes discrete, more precisely, quasidiscrete (see Fig. 5). Fermions in edge states have a finite lifetime because of the nonconservation of the tangential momentum component and displacement of Dirac fermions from the edge to the continuum of bulk states.

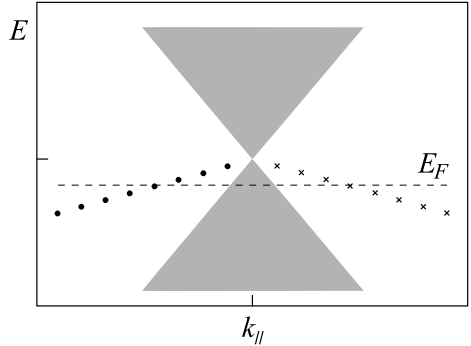

**Fig. 5.** Energy of the Dirac fermion versus the tangential component of the momentum $k_\parallel$ in graphene with a hole of diameter D in the joint valley scheme. The semiclassical orbital quantization $k_\parallel = 2N/D$ with the experimentally extracted sign of the edge parameter $a$ results in the appearance of a ladder of hole type edge levels (see Eq. (2)). Edge Dirac fermions occupying this ladder rotate around the hole clockwise or counterclockwise in the (circles) left or (crosses) right valleys, respectively. The shaded region is the continuum of bulk states.

Neglecting the decay of edge states, we can obtain the following simple expression for the $N$th energy level of the edge state doubly degenerate in the number of the valley and doubly degenerate in the real spin:

$$E_N = 4 \hbar a v_F N/D, \qquad (2)$$

Here, we use the semiclassical orbital quantization condition $\pi D = 2\pi N/k_\parallel$ and $N = 1, 2, \ldots$. In this approximation, spectrum (2) is equidistant. The relation between the position of the Fermi level in graphene and the carrier density $n = V_g \varepsilon_0 \varepsilon / ed$ has the form $E_F = \hbar v_F (\pi n)^{1/2}$ known for massless Dirac fermions. Here, $d$ and $\varepsilon$ are the thickness and relative permittivity of the silicon oxide layer on which graphene is deposited. The resulting expression for the gate voltage $V_{gN}$ corresponding to the resonance condition $E_F = E_N$ has the form

$$V_{gN} = (16 a^2 ed / \pi \varepsilon_0 \varepsilon) (N/D)^2 . \qquad (3)$$

Nanoperforation can be considered as the introduction of additional scatterers for bulk Dirac fermions. It is reasonable to assume that the condition $E_F = E_N$ is accompanied by the resonant scattering of carriers on nanoholes, which results in peaks in the resistance of the sample.

Expression (3) is in agreement with the experiment. The positions of peaks $V_{gN}$ is indeed proportional to $N^2$. Furthermore, the slope of the straight line $V_{gN}(N^2)$ in Fig. 5 should be inversely proportional to $D^2$ under the assumption that the parameter $a$ is the same for samples with different diameters. This property is also in agreement with the experiment within the error. Indeed, for two samples with the diameters of nanoholes of 10 and 2 nm, the slopes of the straight lines $V_{gN}(N^2)$ differ by a factor of 20, whereas the ratio of the squares of the diameters is $25 \pm 30\%$.

The parameter $a$ can be determined from the slope of the $V_{gN}(N^2)$ dependence by comparing with Eq. (3). It appeared to be $\approx 0.07$ with an accuracy of 30%, which is determined by the accuracy of the measurement of the diameter $D$. This value is in quantitative agreement with a value obtained from the magnetic oscillations of the resistance on nanoperforated thin graphene samples [14]. Since the series of peaks is observed on the hole part of the $R(V_g)$ curve, the parameter $a$ is negative. Therefore, edge Dirac fermions are apparently holes.

The energy of the first level $E_1$ and the distance between the levels for samples with a hole diameter of 10 nm are 17.5 mV. The peaks are smeared with an increase in the temperature and disappear at a temperature of about 60 K corresponding to the condition $E_1 \sim 3kT$. The peaks are also smeared with an increase in the measuring current at low temperatures. This is observed when the lateral voltage on the sample at high currents becomes comparable with $E_1/e$.

## 5. CONCLUSIONS

To summarize, a series of regular peaks located asymmetrically with respect to the Dirac point has been revealed on the gate voltage dependence of the resistance of perforated graphene at low temperatures. The position of the $N$th peak $V_{gN}$ measured from the voltage corresponding to the Dirac point is proportional to $N^2$. The velocity of edge carriers determined from the slope of this straight line is an order of magnitude lower than the velocity of bulk Dirac fermions. The effect is attributed to the quantization of orbital motion of edge Dirac fermions around a nanohole. As a result, an equidistant ladder of quasidiscrete levels of edge Dirac fermions is formed on each nanohole. When the gate voltage is varied, the Fermi level successively intersects the levels of this ladder; as a result, the peaks of the resistance appear. The only phenomenological parameter of the theory $a \approx -0.07$ has been obtained by comparing the theory of edge states in graphene [5].

This work was supported by the Russian Foundation for Basic Research (project nos. 11-02-01379a, 11-02-01290-a, and 11-02-12167-ofi-m-2011), by the Ministry of Education and Science of the Russian Federation (contract no. 8033), and jointly by the Russian Academy of Sciences and Seventh Framework Program "Transnational Access," European Union (contract no. 228043-Euromagnet II-Integrated Activities).

**REFERENCES**

1. K. S. Novoselov, A. K. Geim, S. V. Morozov, et al., Science **306**, 666 (2004).

2. K. S. Novoselov, A. K. Geim, S. V. Morozov, et al., Nature **438**, 197 (2005).

3. K. Nakada, M. Fujita, and M. S. Dresselhaus, Phys.Rev. B **54**, 17954 (1996).

4. A. R. Akhmerov and C. W. J. Beenakker, Phys. Rev. B **77**, 085423 (2008).

5. V. A. Volkov and I. V. Zagorodnev, Low Temp. Phys. **35**, 2 (2009).

6. Yu. I. Latyshev, A. Yu. Latyshev, A. P. Orlov, et al., JETP Lett. **90**, 480 (2009).

7. Yu. I. Latyshev, A. P. Orlov, E. G. Shustin, et al., J. Phys.: Conf. Ser. **248**, 012001 (2010).

8. V. A. Skuratov, S. J. Zinkle, A. E. Efimov, et al., Nucl.Instrum. Methods Phys. Res. B **203**, 136 (2003).

9. V. A. Volkov and T. N. Pinsker, Sov. Phys. Solid State **23**, 1022 (1981).

10. B. A. Volkov, B. G. Idlis, and M. Sh. Usmanov, Phys. Usp. **38**, 761 (1995).

11. G. Tkachov and M. Hentschel, Eur. Phys. J. B **69**, 499 (2009).

12. J. A. M. van Ostaay, A. R. Akhmerov, C. W. J. Beenakker, and M. Wimmer, Phys. Rev. B **84**, 195434 (2011).

13. B. A. Volkov and O. A. Pankratov, JETP Lett. **42**, 178 (1985).

14. Y. Latyshev, A. Orlov, V. Volkov, and P. Monceau, in *Proceedings of the International Conference of High Magnetic Fields in Semiconductor Physics, HMF-20, Chamonix Mont-Blans, France, July 22–27, 2012*.